\documentclass[12pt]{article}
\usepackage{epsf}
\title{ {\bf
$t\rightarrow c H^0$ decay in the general two Higgs doublet model.}}
\author{\vspace{1cm}\\
        {\bf E. O. Iltan}
        \thanks{E-mail address:
        eiltan@heraklit.physics.metu.edu.tr}
 \\
        Physics Department, Middle East Technical University \\
        Ankara, Turkey\\}

\date{}

\begin{document}
\setlength{\baselineskip}{24pt}
\maketitle
\setlength{\baselineskip}{7mm}
\begin{abstract}
We study the flavor changing $t\rightarrow c H^0$ decay in the framework of 
the general two Higgs doublet model, so called model III. Here, we take the 
Yukawa couplings complex and switch on the CP violating effects. We predict
the branching ratio  six orders larger compared to the one calculated in the 
SM, namely $BR \sim 10^{-7}$, and observe a measurable CP asymmetry, at the 
order of $\sim 10^{-2}$.
\end{abstract} 
\thispagestyle{empty}
\newpage
\setcounter{page}{1}
%%%
%%%
%%
\section{Introduction}
The top quark reaches great interest since it breaks the $SU(2)\times U(1)$ 
symmetry maximally and it has rich decay products due to its large mass. The
rare decays of the top quark have been studied in the literature, in the
framework of the standard model (SM) and beyond \cite{Mahlon}-\cite{Tao}; 
the one-loop flavor changing transitions $t\rightarrow c g(\gamma, Z)$ in 
\cite{Peccei,Eilam3} and $t\rightarrow c H^0$ in \cite{Eilam, Eilam3, 
Mele, Tao}. 

In the SM, these decays are suppressed as a result of Glashow-Iliopoulos-
Miaiani (GIM) mechanism \cite{Glashow}. The branching ratios ($BR$) of the 
decays $t\rightarrow c g(\gamma, Z)$ have been predicted in the SM as 
$4 \times 10^{-11}\, (5 \times 10^{-13},\, 1.3 \times 10^{-13}\,)$ in 
\cite{Eilam}. $t\rightarrow c H^0$, which can give strong clues about the 
nature of electroweak symmetry breaking, has been calculated at the order 
of the magnitude of $10^{-14}-10^{-13}$ in the SM, in \cite{Mele}. These 
are small rates for the measurement, even at the highest luminosity 
accelerators and therefore, there is a need to analyse these rare decays in 
a new physics beyond the SM.  

In the present work, we study the flavor changing $t\rightarrow c H^0$ decay
in the framework of the general two Higgs doublet model (model III),where 
$H^0$ is the SM Higgs boson. Here, we take the Yukawa couplings complex and 
switch on the CP violating effects. Since the $BR$ is at the order of 
$\sim 10^{-13}$ in the SM, we neglect this contribution and calculate the 
new physics effects in the model III. In the calculations, we take into 
account the interactions due to the internal mediating charged Higgs boson 
$H^{\pm}$ and neglect the ones including internal neutral Higgs bosons, 
$h^0$  and $A^0$ (see Discussion part). The numerical results show that the 
$BR$ of this process can reach to the values of order $10^{-6}$, playing 
with the free parameters of the model III, respecting the existing 
experimental restrictions. This prediction is almost seven orders larger 
compared to the one in the SM and it is a measurable quantity in the 
accelerators. 

Furthermore, we predict the possible CP asymmetry $A_{CP}$ at the order
of $10^{-2}$ for the intermediate values of the CP parameter, due to the
complex Yukawa coupling $\xi^D_{N,bb}$ (see section 2 for its definition). 
This is purely a new physics effect and the measurement of $A_{CP}$ for the
process under consideration may open a new window to go beyond the SM and
test the new physics.        

The paper is organized as follows:
In Section 2, we present the $BR$ and $A_{CP}$ of the decay $t\rightarrow c
H^0$ in the framework of model III. Section 3 is devoted to discussion and 
our conclusions.

%%%
%%%
\section{The Flavor changing $t\rightarrow c H^0$ decay in the framework of
the general two  Higgs Doublet model} 
The flavor changing transition $t\rightarrow c H^0$ is quite suppressed in
the SM due to the Glashow-Iliopoulos-Maiani (GIM) mechanism \cite{Glashow}. 
The extended Higgs sector could bring large contributions to this decay and 
make CP violation possible, in general. This section is devoted to the 
calculation of the $BR$ and the CP violating asymmetry of the decay under
consideration, in the general two Higgs doublet model, so called model III. 
In the model III, the flavor changing neutral currents in the tree
level are permitted and various new parameters, such as Yukawa couplings,
masses of new Higgs bosons, exist.

The $t\rightarrow c H^0$ process is controlled by the Yukawa interaction and 
in the model III, it reads as  
\begin{eqnarray}
{\cal{L}}_{Y}=\eta^{U}_{ij} \bar{Q}_{i L} \tilde{\phi_{1}} U_{j R}+
\eta^{D}_{ij} \bar{Q}_{i L} \phi_{1} D_{j R}+
\xi^{U\,\dagger}_{ij} \bar{Q}_{i L} \tilde{\phi_{2}} U_{j R}+
\xi^{D}_{ij} \bar{Q}_{i L} \phi_{2} D_{j R} + h.c. \,\,\, ,
\label{lagrangian}
\end{eqnarray}
where $L$ and $R$ denote chiral projections $L(R)=1/2(1\mp \gamma_5)$,
$\phi_{i}$ for $i=1,2$, are the two scalar doublets, 
$\bar{Q}_{i L}$ are left handed quark doublets, $U_{j R} (D_{j R})$ are 
right handed up (down) quark singlets, with  family indices $i,j$. The 
Yukawa matrices $\eta^{U,D}_{ij}$ and $\xi^{U,D}_{ij}$ have in general 
complex entries. It is possible to collect SM particles in the first 
doublet and new particles in the second one by choosing the 
parametrization for $\phi_{1}$ and $\phi_{2}$ as
\begin{eqnarray}
\phi_{1}=\frac{1}{\sqrt{2}}\left[\left(\begin{array}{c c} 
0\\v+H^{0}\end{array}\right)\; + \left(\begin{array}{c c} 
\sqrt{2} \chi^{+}\\ i \chi^{0}\end{array}\right) \right]\, ; 
\phi_{2}=\frac{1}{\sqrt{2}}\left(\begin{array}{c c} 
\sqrt{2} H^{+}\\ H_1+i H_2 \end{array}\right) \,\, .
\label{choice}
\end{eqnarray}
with the vacuum expectation values,  
\begin{eqnarray}
<\phi_{1}>=\frac{1}{\sqrt{2}}\left(\begin{array}{c c} 
0\\v\end{array}\right) \,  \, ; 
<\phi_{2}>=0 \,\, ,
\label{choice2}
\end{eqnarray}
Here, $H_1$ and $H_2$ are the mass eigenstates $h^0$ and $A^0$ respectively 
since no mixing occurs between two CP-even neutral bosons $H^0$ and $h^0$ 
at tree level, for our choice. 

The Flavor Changing (FC) interaction can be obtained as 
\begin{eqnarray}
{\cal{L}}_{Y,FC}=
\xi^{U\,\dagger}_{ij} \bar{Q}_{i L} \tilde{\phi_{2}} U_{j R}+
\xi^{D}_{ij} \bar{Q}_{i L} \phi_{2} D_{j R} + h.c. \,\, ,
\label{lagrangianFC}
\end{eqnarray}
where the couplings  $\xi^{U,D}$ for the FC charged interactions are
\begin{eqnarray}
\xi^{U}_{ch}&=& \xi^U_{N} \,\, V_{CKM} \nonumber \,\, ,\\
\xi^{D}_{ch}&=& V_{CKM} \,\, \xi^D_{N} \,\, ,
\label{ksi1} 
\end{eqnarray}
and $\xi^{U,D}_{N}$ is defined by the expression 
\begin{eqnarray}
\xi^{U (D)}_{N}=(V_{R (L)}^{U (D)})^{-1} \xi^{U,(D)} V_{L(R)}^{U (D)}\,\, .
\label{ksineut}
\end{eqnarray}
Here the index "N" in $\xi^{U,D}_{N}$ denotes the word "neutral". 

\begin{figure}[htb]
\vskip -3.0truein
\centering
\epsfxsize=6.8in
\leavevmode\epsffile{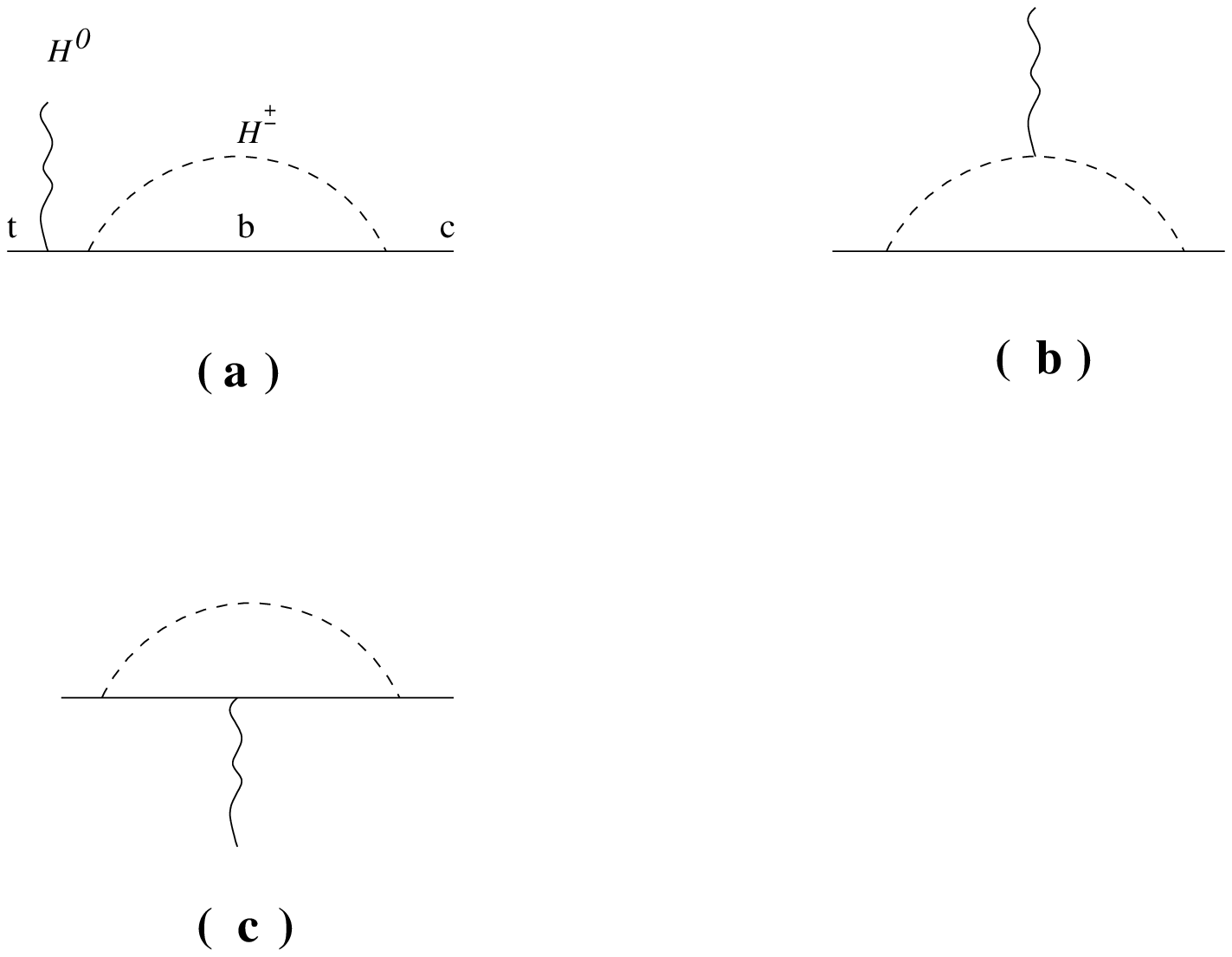}
\vskip -2.0truein
\caption[]{One loop diagrams contribute to the decay $t\rightarrow c H^0$ 
due to internal charged Higgs boson. Wavy line represents the $H^0$ 
field and dashed line the $H^{\pm}$ field.}
\label{fig1}
\end{figure}

The SM contribution to the $BR$ of the process  $t\rightarrow c H^0$ is 
negligibly small, which is at the order of the magnitude $10^{-13}$.
Therefore, we take into account only the new effects beyond the SM. The 
relevant diagrams are given in Fig \ref{fig1}. At this stage, we would like
to discuss the possibilities not to take into account the tree level 
contribution to the decay underconsideration, in the model III. First, it 
can be assumed that all the off diagonal neutral Yukawa couplings vanish 
and therefore the coupling $\xi^{U}_{N,tc}$ vanishes. Another possibility is 
to take the mixing between two neutral Higgs bosons $H^0$ and $h^0$ is
small and the tree level interaction $t-c-H^0$ is negligible. In our case, we 
consider the gauge and $CP$ invariant Higgs potential which 
spontaneously breaks  $SU(2)\times U(1)$ down to $U(1)$  as:
\begin{eqnarray}
V(\phi_1, \phi_2 )&=&c_1 (\phi_1^+ \phi_1-v^2/2)^2+
c_2 (\phi_2^+ \phi_2)^2 \nonumber \\ &+& +
c_3 [(\phi_1^+ \phi_1-v^2/2)+ \phi_2^+ \phi_2]^2
+ c_4 [(\phi_1^+ \phi_1) (\phi_2^+ \phi_2)-(\phi_1^+ \phi_2)(\phi_2^+ \phi_1)]
\nonumber \\ &+& 
c_5 [Re(\phi_1^+ \phi_2)]^2 +
c_{6} [Im(\phi_1^+ \phi_2)]^2 
+c_{7}\,\, .
\label{potential}
\end{eqnarray}
Since we assume that only $\phi_1$ has vacuum expectation value, no mixing
occurs between two neutral Higgs bosons and the tree level interaction
vanishes. Furthermore, since we take the c-quark mass zero and the coupling 
$\xi^{U}_{N,tc}$ negligible compared to the couplings $\xi^{U}_{N,tt}$ and 
$\xi^{D}_{N,bb}$ (see \cite{Alil1}), the main contribution comes  from the 
diagrams with internal charged Higgs boson and for the matrix element, we 
get 
\begin{eqnarray}
M(t\rightarrow c H^0)= -i V_{cb}\, V^*_{tb}\frac{g}{64\, m_W\, \pi^2} 
(F^{(vert)}+F^{(self)})\,\bar{c}\,(1+\gamma_5)\,t \,\, ,
\label{M1}
\end{eqnarray} 
where 
\begin{eqnarray}
F^{(vert)}&=& \xi^D_{N,bb}\, (m_b\, \xi^{U \,*}_{N,tt}\,f_1+m_t\,
\xi^{D \,*}_{N,bb}\, f_2)\nonumber \,\, , \\
F^{(self)}&=& m_b\, \xi^D_{N,bb}\,\xi^{U \,*}_{N,tt}\, f_3\,\, ,
\label{Fverself}
\end{eqnarray}
and the functions $f_1$, $f_2$, $f_3$ are defined as 
\begin{eqnarray}
f_1&=& \int_{0}^{1} dx\,\int_{0}^{1-x} dy\, \Bigg\{ 1+\frac{z_t\,(-1+x+y)\,
(y+x\,zt)} {x+g\,(x,y)}+\frac{y_W\,(2-\frac{1}{cos^2\,\theta_W})}{1-x+g\,
(x,y)}+2\,ln\, ( x+g\,(x,y) ) \Bigg \}\nonumber \,\, , \\ 
f_2&=& \int_{0}^{1} dx\,\int_{0}^{1-x} dy\,\frac{y_W\,(-1+x+y)\, 
(2-\frac{1}{cos^2\,\theta_W})}{1-x+g\,(x,y)} 
\nonumber \,\, , \\ 
f_3&=& \frac{1-y_b+y_b\, ln y_b}{1-y_b} \,\, ,
\label{f1f2f3}
\end{eqnarray}
with  $g\,(x,y)=_(-1+x+y)\,(x\,y_t+y\,z_t)$. Here 
$y_t=m_t^2/m^2_{H^{\pm}},\, z_t=m^2_{H^0}/m^2_{H^{\pm}},\, 
y_W=m_W^2/m^2_{H^{\pm}}$ and $y_b=m_b^2/m^2_{H^{\pm}}$. Using the eq. 
(\ref{M1}), it is straightforward to obtain the decay width as 
\begin{eqnarray}
\Gamma (t\rightarrow c H^0) = \frac{1}{32\,\pi} \frac{y_t-z_t}{m_t\, y_t}
\, |M|^2 \,\, .
\label{DW}
\end{eqnarray}

Now we would like to give the expression for $A_{CP}$ of the above process. 
Here, we take $\xi^D_{N,bb}$ and $\xi^{U}_{N,tt}$ complex with the 
parametrizations 
\begin{eqnarray}
\xi^{U}_{N,tt}=|\xi^{U}_{N,tt}|\, e^{i \theta_{tt}}
\nonumber \, , \\
\xi^{D}_{N,bb}=|\xi^{D}_{N,bb}|\, e^{i \theta_{tb}} \, ,
\label{xi}
\end{eqnarray}
and assume that the complexity of $\xi^{U}_{N,tt}$ is small. 
Using the definition of the CP violating asymmetry $A_{CP}$ 
\begin{eqnarray}
A_{CP}= \frac{\Gamma (t\rightarrow c H^0) -\Gamma (\bar{t}\rightarrow
\bar{c} H^0)}{\Gamma (t\rightarrow c H^0) +\Gamma (\bar{t}\rightarrow
\bar{c} H^0)}
\label{cpvio}
\end{eqnarray}
we get 
\begin{eqnarray}
A_{CP}=|\xi^{U}_{N,tt}|\, 
sin(\theta_{tb}-\theta_{tt})\, \frac{N}{D} \,\, ,
\label{Acp}
\end{eqnarray}
where
\begin{eqnarray}
N&=& -2\, Im \Big( f_1\,f_2^* +f_3\,f_2 \Big)\nonumber \,\, , \\
D&=& \frac{m_t}{m_b}\,y_W\, |\xi^{D}_{N,bb}|\,|f_2|^2+ 2\, |\xi^{U}_{N,tt}|
\, Re \Big( f_1\,f_2^* -f_3\,f_2 \Big) cos(\theta_{tb}-\theta_{tt})\,\, ,
\label{NumDen}
\end{eqnarray}
and the functions $f_1$, $f_2$ and $f_3$ are given in eq. (\ref{f1f2f3}). 
Here the symbol $*$ denotes the complex conjugation. As it is shown from 
the eq. (\ref{Acp}), $A_{CP}$ vanishes when two CP violating angles 
$\theta_{tb}$ and $\theta_{tt}$ are equal. This is interesting, since 
$A_{CP}$ can vanish even in the existence of complex phases, in the model 
III.
%%%
%%%
\section{Discussion}
In this section, we study the $sin\,\theta_{tb}$, $m_{H^0}$ and 
$\bar{\xi}^{D}_{N,bb}$ dependencies of the $BR$ and $A_{CP}$ for the process 
$t\rightarrow c H^0$, in the model III. For the calculation of the $BR$ we 
take the value of the total decay width $\Gamma_T \sim \Gamma 
(t\rightarrow b W)$ as $\Gamma_T=1.55\, GeV $. Notice that the coupling 
$\bar{\xi}^{D}_{N,bb}$ is defined as $\xi^{U(D)}_{N,ij}=\sqrt{\frac{4\,G_F}
{\sqrt{2}}}\, \bar{\xi}^{U(D)}_{N,ij}$. 

The process $t\rightarrow c H^0$ exists also in the SM and model II (or I)
version of the 2HDM. In both models this process appears at least at loop
level. In the SM model internal mediating $W^{\pm}$ bosons are responsible
for this decay and its BR is very small, at the order of the magnitude of 
$10^{-13}$. In the model II, additional contribution comes from the charged
Higgs boson $H^{\pm}$ and can be enhanced compared to the SM result by
playing with the free parameter $tan\beta$, respecting the experimental
measurements. However, no $A_{CP}$ occurs in the model II (I) and also in
the SM. This forces one to go into the model III, with complex and possibly 
large Yukawa couplings.

Model III contains large number of free parameters such as Yukawa couplings, 
$\bar{\xi}^{U (D)}_{N, ij}$, the masses of new Higgs bosons, $H^{\pm}$, 
$h^0$ and $A^0$, and they should be restricted by using experimental 
measurements. At this stage we summarize these restrictions:
\begin{itemize}
\item We neglect all the Yukawa couplings except $\bar{\xi}^{U}_{N, tt}$ and 
$\bar{\xi}^{D}_{N, bb}$ since they are negligible due to their light flavor 
contents, by our assumption. Notice that we also neglect the off diogonal 
coupling $\bar{\xi}^{U}_{N, tc}$, since it is smaller compared to 
$\bar{\xi}^{U}_{N,tt}$ (see \cite{Alil1}). Therefore,  the new neutral Higgs 
bosons do not have any contribution to the $BR$ of the decay under 
consideration.

\item We take $\bar{\xi}^{U}_{N, tt}$ and $\bar{\xi}^{D}_{N, bb}$ complex 
and respect the constraint for the angle $\theta_{tt}$ and $\theta_{bb}$, 
due to the experimental upper limit of neutron electric dipole moment, 
$d_n<10^{-25}\hbox{e$\cdot$cm}$, which leads to $\frac{1}{m_t m_b} 
Im(\bar{\xi}^{U}_{N, tt}\, \bar{\xi}^{* D}_{N, bb})< 1.0$ for $M_{H^\pm}
\approx 200$ GeV \cite{david}. 

\item We find a constraint region for these free parameters by restricting 
the Wilson coefficient $C_7^{eff}$, which is the effective coefficient of 
the operator $O_7 = \frac{e}{16 \pi^2} \bar{s}_{\alpha} 
\sigma_{\mu \nu} (m_b R + m_s L) b_{\alpha} {\cal{F}}^{\mu \nu}$
(see \cite{Alil1} and references therein), in the region 
$0.257 \leq |C_7^{eff}| \leq 0.439$. Here upper and lower limits were 
calculated using the CLEO measurement \cite{cleo2}
\begin{eqnarray}
BR (B\rightarrow X_s\gamma)= (3.15\pm 0.35\pm 0.32)\, 10^{-4} \,\, .
\label{br2}
\end{eqnarray}
and all possible uncertainities in the calculation of $C_7^{eff}$ 
\cite{Alil1}. 
\end{itemize}

For completeness we present the Wilson coefficient $C_7^{eff}$. 
Denoting the contribution for the SM with $C_{7}^{SM}(m_{W})$ and 
the additional charged Higgs contribution with $C_{7}^{H}(m_{W})$, 
we have the initial values  
\begin{eqnarray}
C_7^{2HDMH}(m_W)&=&C_{7}^{SM}(m_{W})+\frac{1}{m_{t}^2} \,
(\bar{\xi}^{* U}_{N,tt}+\bar{\xi}^{* U}_{N,tc}
\frac{V_{cs}^{*}}{V_{ts}^{*}}) \, (\bar{\xi}^{U}_{N,tt}+\bar{\xi}^{U}_{N,tc}
\frac{V_{cb}}{V_{tb}}) F_{1}(y_t)\nonumber  \, \, , \\
&+&\frac{1}{m_t m_b} \, (\bar{\xi}^{* U}_{N,tt}+\bar{\xi}^{* U}_{N,tc}
\frac{V_{cs}^{*}}{V_{ts}^{*}}) \, (\bar{\xi}^{D}_{N,bb}+\bar{\xi}^{D}_{N,sb}
\frac{V_{ts}}{V_{tb}}) F_{2}(y_t)
\, \, . 
\label{CoeffH}
\end{eqnarray}

The LO corrected coefficient $C_{7}^{2HDM}(\mu)$ is given as 
\begin{eqnarray} 
C_{7}^{LO, 2HDM}(\mu)&=& \eta^{16/23} C_{7}^{2HDM}(m_{W})+(8/3) 
(\eta^{14/23}-\eta^{16/23}) C_{8}^{2HDM}(m_{W})\nonumber \,\, \\
&+& C_{2}^{2HDM}(m_{W}) \sum_{i=1}^{8} h_{i} \eta^{a_{i}} \,\, , 
\label{LOwils}
\end{eqnarray}
and $\eta =\alpha_{s}(m_{W})/\alpha_{s}(\mu)$, $h_{i}$ and $a_{i}$ are 
the numbers which appear during the evaluation \cite{buras}. 
The explicit forms of the functions $F_{1(2)}(y)$ are, 
\begin{eqnarray}
F_{1}(y)&=& \frac{y(7-5y-8y^2)}{72 (y-1)^3}+\frac{y^2 (3y-2)}{12(y-1)^4}
\,\ln y \nonumber  \,\, , \\ 
F_{2}(y)&=& \frac{y(5y-3)}{12 (y-1)^2}+\frac{y(-3y+2)}{6(y-1)^3}\, \ln y 
\nonumber  \,\, .
\label{F1G1}
\end{eqnarray}

The discussion given above allows us to obtain a constraint region for the
couplings $\bar{\xi}^{U}_{N, tt}$, $\bar{\xi}^{D}_{N, bb}$ and the CP 
violating parameters, $sin\,\theta_{tt}$ and $sin\,\theta_{tb}$. Here, we 
assume that the coupling $\bar{\xi}^{U}_{N, tt}$ has a small imaginary part 
and we choose $|r_{tb}|=|\frac{\bar{\xi}_{N, tt}^{U}}{\bar{\xi}_{N, bb}^{D}}| 
<1$. Notice that, in figures,  the $BR$ is restricted in the region between 
solid (dashed) lines for $C_7^{eff} > 0$ ($C_7^{eff} < 0$). Here, 
there are two possible solutions for $C_7^{eff}$ due to the cases where 
$|r_{tb}|<1$ and $r_{tb}>1$. In the case of complex Yukawa couplings, only 
the solutions obeying $|r_{tb}|<1$ exist.

\begin{figure}[htb]
\vskip -3.0truein
\centering
\epsfxsize=6.8in
\leavevmode\epsffile{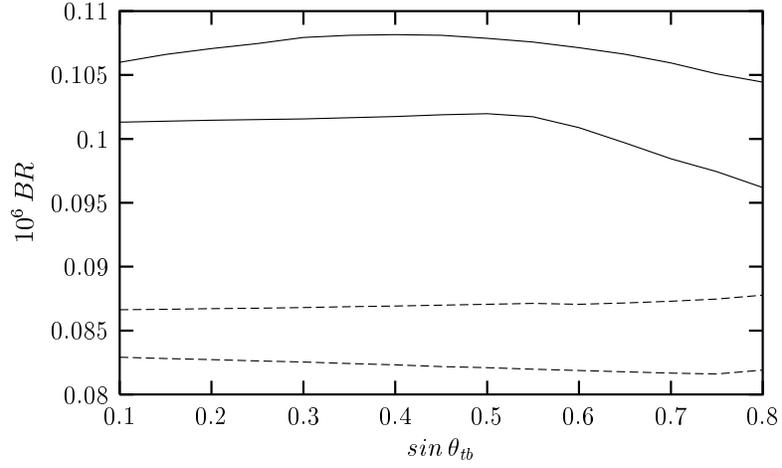}
\vskip -3.0truein
\caption[]{$BR (t\rightarrow c H^0)$ as a function of  $sin\,\theta_{tb}$ 
for $|\bar{\xi}^D_{N,bb}|=10\,m_b$, $m_{H^{\pm}}=400\, GeV$, 
$m_{H^0}=120\, GeV$, $sin\,\theta_{tt}=0.1$ in the model III. Here the $BR$ 
is restricted in the region bounded by solid lines for $C_7^{eff}>0$ and by 
dashed  lines for $C_7^{eff}<0$}
\label{BRsintb}
\end{figure}

In  Fig. \ref{BRsintb}, we plot the $BR$ with respect to $sin\,\theta_{tb}$ 
for $sin\,\theta_{tt}=0.1$, $m_{H^{\pm}}=400\, GeV$, $m_{H^0}=120\, GeV$, 
$\bar{\xi}_{N, bb}^{D}=10\, m_b$.  As shown in this figure, the $BR$ can 
reach to the values at the order of the magnitude of $10^{-7 }$ and it is 
not so much sensitive to the parameter $sin\,\theta_{tb}$. Its magnitude is 
almost $30\%$ larger for $C_7^{eff} > 0$ compared to the one for 
$C_7^{eff} < 0$. 

\begin{figure}[htb]
\vskip -3.0truein
\centering
\epsfxsize=6.8in
\leavevmode\epsffile{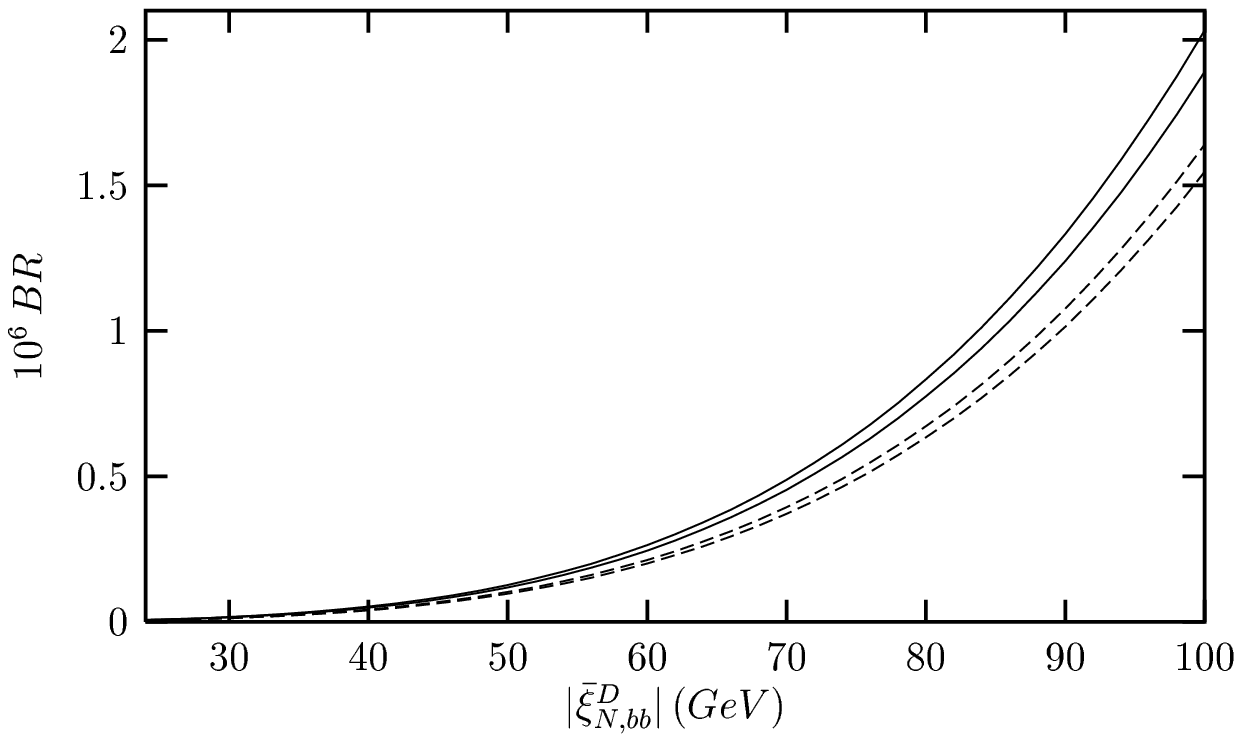}
\vskip -3.0truein
\caption[]{$BR (t\rightarrow c H^0)$ as a function of $|\bar{\xi}^D_{N,bb}|$ 
for $sin\,\theta_{tt}=0.1$, $sin\,\theta_{tb}=0.5$, $m_{H^{\pm}}=400\, GeV$, 
$m_{H^0}=120\, GeV$ in the model III. Here the $BR$ is restricted in the 
region bounded by solid lines for $C_7^{eff}>0$ and by dashed  lines for 
$C_7^{eff}<0$}
\label{BRksibb}
\end{figure}

Fig. \ref{BRksibb} represents the $BR$ with respect to 
$|\bar{\xi}^{D}_{N, bb}|$ for $sin\,\theta_{tt}=0.1$, $sin\,\theta_{tb}=0.5$, 
$m_{H^{\pm}}=400\, GeV$ and $m_{H^0}=120\, GeV$. This figure shows that the 
$BR$ is strongly sensitive to the the coupling $|\bar{\xi}^{D}_{N, bb}|$ and 
it can get the values at the order of $10^{-6}$ even for 
$|\bar{\xi}^{D}_{N, bb}|=20\, m_b$. This observation is an important clue 
about the upper limit of the coupling $|\bar{\xi}^{D}_{N, bb}|$, with the
possible future measurement of the $BR$. For the small coupling 
$|\bar{\xi}^{D}_{N, bb}|$, the restricted region for the $BR$ becomes narrow, 
for both $C_7^{eff} > 0$ and $C_7^{eff} < 0$.

\begin{figure}[htb]
\vskip -3.0truein
\centering
 \epsfxsize=6.8in
\leavevmode\epsffile{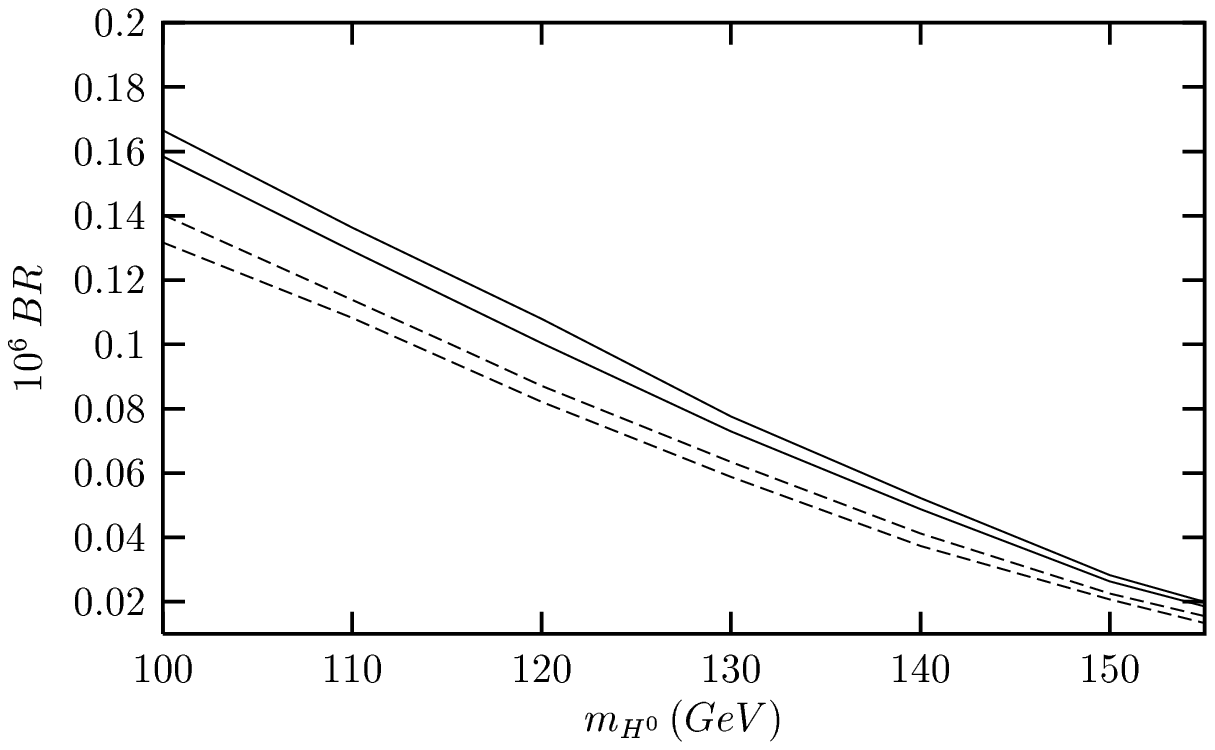}
\vskip -3.0truein
\caption[]{$BR (t\rightarrow c H^0)$ as a function of $m_{H^0}$ 
for $|\bar{\xi}^D_{N,bb}|=10\,m_b$, $sin\,\theta_{tt}=0.1$, 
$sin\,\theta_{tb}=0.5$, $m_{H^{\pm}}=400\, GeV$ in the model III. 
Here the $BR$ is restricted in the region bounded by solid lines for 
$C_7^{eff}>0$ and by dashed  lines for $C_7^{eff}<0$}
\label{BRmH0}
\end{figure}

Finally, we show the $m_{H^0}$ dependence of the $BR$ in Fig. \ref{BRmH0} 
for $sin\,\theta_{tt}=0.1$, $sin\,\theta_{tb}=0.5$, $m_{H^{\pm}}=400\, GeV$ 
and $|\bar{\xi}^{D}_{N, bb}|=10\,m_b$. With the increasing values 
of $m_{H^0}$, the $BR$ decreases and the restriction region becomes narrower.

\begin{figure}[htb]
\vskip -3.0truein
\centering
\epsfxsize=6.8in
\leavevmode\epsffile{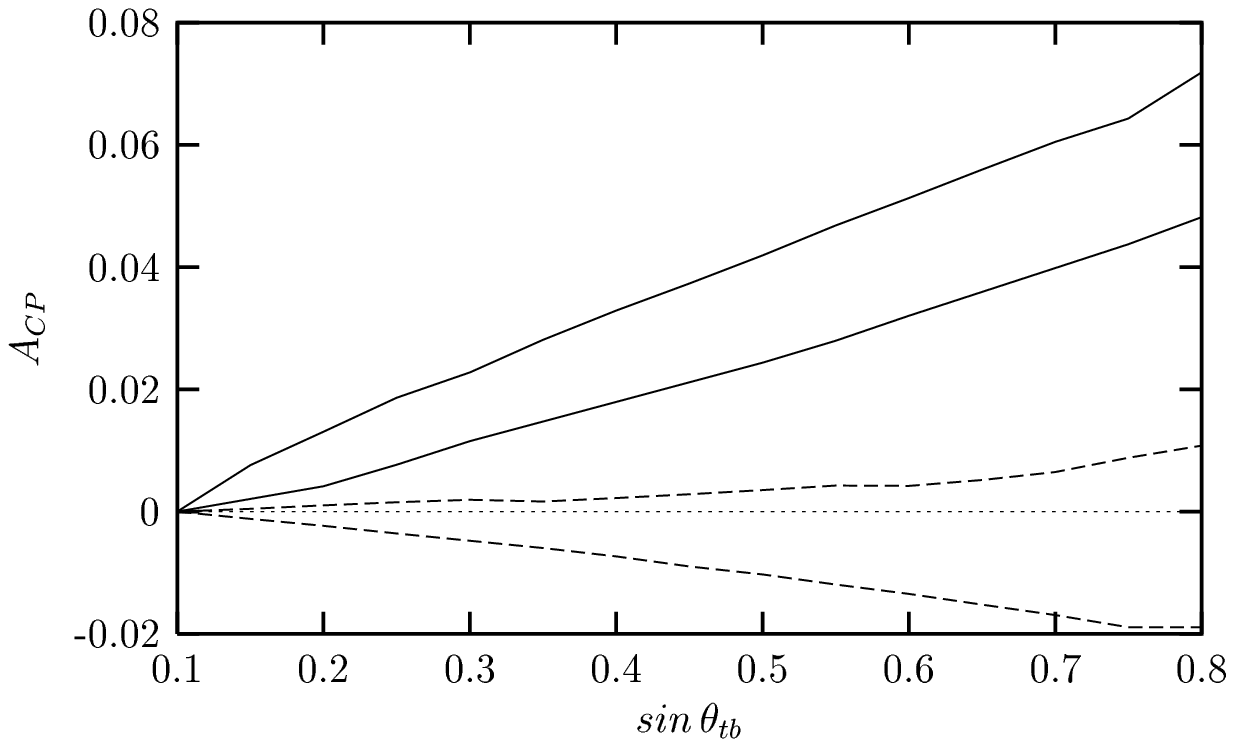}
\vskip -3.0truein
\caption[]{The same as Fig. \ref{BRsintb}, but for $A_{CP}$.}
\label{ACPsintb}
\end{figure}

\begin{figure}[htb]
\vskip -3.0truein
\centering
\epsfxsize=6.8in
\leavevmode\epsffile{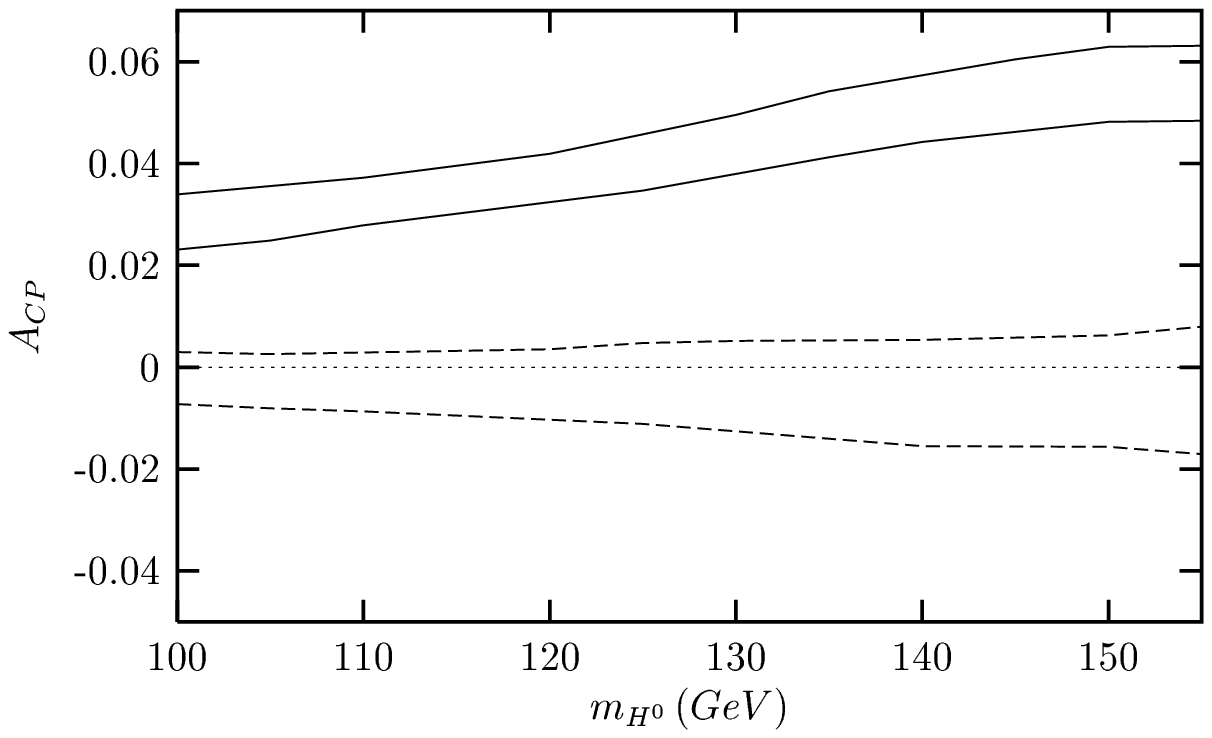}
\vskip -3.0truein
\caption[]{The same as Fig. \ref{BRmH0}, but for $A_{CP}$.}
\label{ACPmH0}
\end{figure}

At this stage, we would like to analyse the CP asymmetry $A_{CP}$ of the decay
$t\rightarrow c H^0$ and show $sin\,\theta_{tb}$ and $m_{H^0}$
dependencies of $A_{CP}$  in the figures \ref{ACPsintb} and 
\ref{ACPmH0}, respectively. $A_{CP}$ is at the order of the magnitude of 
$10^{-2}$ for the intermeditate values of $sin\,\theta_{tb}$ and it can 
reach to $7\times 10^{-2}$ for $C_7^{eff} > 0$. Notice that 
$sin\,\theta_{tt}$ is taken small and $A_{CP}$ vanishes when two CP 
parameters have the same values, namely $sin\,\theta_{tb}=sin\,\theta_{tt}$. 
If $A_{CP}$ is positive (negative), $C_{7}^{eff}$ can have both signs. 
However, if it is negative, $C_{7}^{eff}$ must be negative. This observation 
is useful in the determination of the sign of $C_{7}^{eff}$. The same 
behaviour is observed in Fig. \ref{ACPmH0} which represents the mass 
$m_{H^0}$ dependence of $A_{CP}$.  When the Higgs mass $m_{H^0}$ increases, 
an enhancement in $A_{CP}$ is detected, especially for $C_{7}^{eff}>0$.  

Now we will summarize our results:

\begin{itemize}

\item The $BR$ of the flavor changing process $t\rightarrow c H^0$ is at the 
order of $10^{-13}$ in the SM and the extended Higgs sector brings large 
contributions, at the order of $10^{-7}-10^{-6}$, which can be measured in 
the future experiments. This ensures a crucial test for the new physics 
beyond the SM. 

\item The $BR$ is sensitive to $|\bar{\xi}^{D}_{N, bb}|$ and its measurement 
makes it possible to predict an upper limit for this coupling. Furthermore, 
the measurement of the $BR$ of the decay under consideration can give 
important information about the mass of Higgs boson $H^0$.    

\item $A_{CP}$ is at the order of the magnitude of $10^{-2}$ for the 
intermeditate values of $sin\,\theta_{tb}$ and it rises up to the values 
$7\times 10^{-2}$ for $C_7^{eff} > 0$. The measurement of $A_{CP}$ can
ensure a hint for the determination the sign of  $C_7^{eff}$.

\end{itemize}

Therefore, the experimental investigation of the process $t\rightarrow c 
H^0$ will be effective for understanding the physics beyond the SM.
\section{Acknowledgement}
This work was supported by Turkish Academy of Sciences (TUBA/GEBIP).

\newpage
\newpage

\begin{thebibliography}{1}
%
\bibitem{Mahlon}
G.Mahlon and S.Parke, {\it Phys. Lett.} {\bf B347} (1995) 394.
%
\bibitem{Jenkins}
E. Jenkins, {\it Phys. Rev.} {\bf D56} (1997) 458; D. Atwood and M. Sher, 
{\it Phys. Lett. } {\bf B411} (1997) 306.
%
\bibitem{Eilam}
G.Eilam, J.L.Hewett and A.Soni, {\it Phys. Rev.} {\bf D44} (1991) 1473, 
erratum {\it Phys. Rev.} {\bf D59} 039901 (1999).
%
\bibitem{Li}
C.S.Li, R.J.Oakes and J.M.Yang, {\it Phys. Rev.} {\bf D49} (1994) 293; 
J.M.Yang and  C.S.Li, {\it Phys. Rev.} {\bf D49} (1994) 3412; G.Couture, 
C.Hamzaoui and H.K\"onig, {\it Phys. Rev.} {\bf D52} (1995) 171; J.L.Lopez, 
D.V.Nanopoulos and R.Rangarajan,  {\it Phys. Rev.} {\bf D56} (1997) 3100; 
G.Couture, M.Frank and H.K\"onig, {\it Phys. Rev.} {\bf D56} (1997) 4213; 
G.M.de Divitiis, R.Petronzio, L.Silvestrini, {\it Nucl. Phys.} {\bf B504} 
(1997) 45.
%
\bibitem{Peccei} 
T.Han, R.D.Peccei, and X.Zhang, {\it Nucl. Phys.} {\bf B454} (1995) 527;
T.Han, K.Whisnant, B.-L.Young, and X.Zhang, {\it Phys. Rev.} {\bf D55} 
(1997) 7241; {\it Phys. Lett.} {\bf B385} (1996) 311;
M.Hosch, K.Whisnant, and  B.-L.Young, {\it Phys. Rev.} {\bf D56} (1997)
5725; E. Malkawi and T. Tait, {\it Phys. Rev.} {\bf D54} (1996) 5758;
T. Tait and C.P. Yuan,  {\it Phys. Rev.} {\bf D55} (1997) 7300;
T. Han, M. Hosch, K. Whisnant, B.-L. Young and X. Zhang, 
{\it Phys. Rev.} {\bf D58} (1998) 073008; F. del Aguila and J.A. 
Aguilar-Saavedra, {\it Phys. Lett.} {\bf B462} (1999) 310; 
F. del Aguila and J.A. Aguilar-Saavedra, {\it Nucl. Phys.} {\bf B576} 
(2000) 56.
%
\bibitem{Hou}
W.S.Hou, {\it Phys. Lett.} {\bf B296} (1992) 179; K.Agashe, M.Graesser, 
{\it Phys. Rev.} {\bf D54} (1996) 4445; M.Hosch, K.Whisnant, B.L.Young, 
{\it Phys. Rev.} {\bf D56} (1997) 5725; J.Guasch, hep-ph/9710267.
%
\bibitem{intox}
L.T.Handoko and J.Hashida, {\it Phys. Rev.} {\bf D58} (1998) 094008.
%
\bibitem{Eilam3}
G.Eilam, B.Haeri and A.Soni, Phys. Rev. D41 (1990) 875.
%
\bibitem{Mele}
B. Mele, S. Petrarca and A. Soddu, {\it Phys. Lett.} {\bf B435}, (1998) 401.
%
\bibitem{Tao}
T. Han and Jing Jiang, B. Mele, {\it Phys. Lett.} {\bf B516}, (2001) 337.
%
\bibitem{Glashow} S. L. Glashow, J. Iliopoulos and L. Maiani, 
{\it Phys. Rev.} {\bf D2} (1970) 875
\bibitem{Alil1} T. M.Aliev, E. Iltan, {\it J. Phys.} {\bf G25}
(1999) 989.
%
\bibitem{david} D. B. Chao, K. Cheung and W. Y. Keung,
{\it Phys. Rev.} {\bf 59} (1999) 115006.
%
\bibitem{cleo2}  S. Chen, et. al., CLEO Collaboration, {\it Phys. Rev. Lett.} 
{\bf 87} (2001) 251807.
%
\bibitem{buras} A. J. Buras and M. M\"{u}nz,
{\it Phys. Rev.} {\bf D52} (1995) 186.
%
\end{thebibliography}
\end{document}